\documentclass[apjl]{emulateapj}
\usepackage{txfonts}
\usepackage[dvipsnames,usenames]{xcolor}
\usepackage{soul}

\newcommand{\code}[1]{\texttt{#1}}

\newcommand*{\mxb}{MXB~1659-29}

\newcommand*{\ks}{KS~1731-260}
\newcommand*{\sgr}{SGR~1627-41}
\newcommand*{\qimp}{\ensuremath{Q_{\rm imp}}}
\newcommand*{\Tcore}{\ensuremath{T_{\rm core}}}
\newcommand*{\Ggap}{\ensuremath{\rm G08}}
\newcommand*{\Sgap}{\ensuremath{\rm S03}}

\newcommand*{\Qimp}{\qimp}

\newcommand*{\K}{\mathrm{K}}

\newcommand*{\mesa}{\code{MESA}}
\newcommand*{\dStar}{\code{dStar}}
\newcommand*{\Zbar}{\ensuremath{\langle Z\rangle}}
\newcommand*{\Fermi}[2]{\ensuremath{#1_{\mathrm{F},#2}}}
\newcommand*{\EFe}{\Fermi{E}{e}}

\newcommand*{\pFn}{\Fermi{p}{n}}
\newcommand*{\pFe}{\Fermi{p}{e}}
\newcommand*{\kFn}{\Fermi{k}{n}}

\newcommand*{\vFn}{\Fermi{v}{n}}
\newcommand*{\vFe}{\Fermi{v}{e}}
\newcommand*{\kB}{\ensuremath{k_{\mathrm{B}}}}
\newcommand*{\mnstar}{\ensuremath{m_n^{\star}}}

\sethlcolor{Goldenrod}
\setstcolor{Red}

\begin{document}
\title{Late time cooling of neutron star transients and the physics of the inner crust}
\author{Alex Deibel\altaffilmark{1,2$^{\star}$}, Andrew Cumming\altaffilmark{2,3}, Edward F. Brown\altaffilmark{1,2,4}, and Sanjay Reddy\altaffilmark{2,5}}
\affil{
\altaffilmark{1}{Department of Physics and Astronomy, Michigan
State University, East Lansing, MI 48824, USA} \\
\altaffilmark{2}{The Joint Institute for Nuclear Astrophysics - Center for the Evolution of the Elements, Michigan State University, East Lansing, MI 48824, USA} \\
\altaffilmark{3}{Department of Physics and McGill Space Institute, McGill University, 3600 rue University, Montreal, QC, H3A 2T8, Canada} \\
\altaffilmark{4}{National Superconducting Cyclotron Laboratory, Michigan State University,
East Lansing, MI 48824, USA} \\
\altaffilmark{5}{Institute for Nuclear Theory, University of Washington, Seattle, WA 98195-1550, USA} \\
}
\shorttitle{LATE TIME COOLING AND THE PHYSICS OF THE INNER CRUST}
\shortauthors{DEIBEL ET AL.}

\email{$^{\star}$deibelal@msu.edu}

\begin{abstract}
An accretion outburst onto a neutron star transient heats the neutron star's crust out of thermal equilibrium with the core. After the outburst the crust thermally relaxes toward equilibrium with the neutron star core and the surface thermal emission powers the quiescent X-ray light curve. Crust cooling models predict that thermal equilibrium of the crust will be established $\approx 1000 \, \mathrm{d}$ into quiescence. Recent observations of the cooling neutron star transient \mxb, however, suggest that the crust did not reach thermal equilibrium with the core on the predicted timescale and continued to cool after $\approx 2500 \, \mathrm{d}$ into quiescence. Because the quiescent light curve reveals successively deeper layers of the crust, the observed late time cooling of \mxb\ depends on the thermal transport in the inner crust. In particular, the observed late time cooling is consistent with a low thermal conductivity layer near the depth predicted for nuclear pasta that maintains a temperature gradient between the neutron star's inner crust and core for thousands of days into quiescence. As a result, the temperature near the crust-core boundary remains above the critical temperature for neutron superfluidity and a layer of normal neutrons forms in the inner crust. We find that the late time cooling of \mxb\ is consistent with heat release from a normal neutron layer near the crust-core boundary with a long thermal time. 



\end{abstract}


\section{Introduction}
\label{sec:intro}

An accretion outburst onto a neutron star transient triggers non-equilibrium nuclear reactions \citep{sato1979,bisnovatyi1979} that deposit heat in the neutron star's crust \citep{haensel1990, haensel2003, haensel2008}. Accretion-driven heating brings the crust out of thermal equilibrium with the core; when accretion ceases, the crust cools toward thermal equilibrium with the core and powers the quiescent light curve \citep{brown1998,ushomirsky2001,rutledge2002}. Brown \& Cumming (2009) discussed the basic idea that observations at successively later times into quiescence probe successively deeper layers in the crust with increasingly longer thermal times. In particular, about a year into quiescence the shape of the cooling light curve is sensitive to the physics of the inner crust at mass densities greater than neutron drip $\rho \gtrsim \rho_{\rm drip} \approx 4 \times 10^{11} \, \mathrm{g \ cm^{-3}}$ \citep{page2012}.



Among the modeled cooling transients, \mxb \ \citep{wijnands2003, wijnands2004, cackett2008} was thought to be unique in that its crust appeared to reestablish its long-term thermal equilibrium with the core after $\approx 1000 \, \mathrm{d}$ into quiescence \citep{brown2009}.  Recent observations of \mxb \ \citep{cackett2013}, however, indicate that the crust continued to cool after $\approx 2500 \, \mathrm{d}$ to reach a new low temperature when observed $\approx 4000 \, \mathrm{d}$ into quiescence. Although the drop in count rate could be explained by a change in absorption column, for example due to a build up of an accretion disk in the binary, it is also consistent with a drop in neutron star effective temperature.

Horowitz et al.~(2015) show that the late time drop in temperature in \mxb \ could be caused by a low thermal conductivity layer at the base of the inner crust at a mass density $\rho \gtrsim 8 \times 10^{13} \, \mathrm{g \ cm^{-3}}$. The low thermal conductivity may be a consequence of nuclear pasta, which forms when nuclei are distorted into various complex shapes at high densities in the inner crust \citep{ravenhall1983,hashimoto1984,oyamatsu1993}. Nuclear pasta has been studied using quantum molecular dynamics simulations \citep{maruyama1998,watanabe2003} and semi-classical molecular dynamics simulations \citep{horowitz2004,horowitz2008, schneider2013}, but the thermal properties of nuclear pasta remain uncertain.  Horowitz et al.~(2015) discovered a possible mechanism for lowering the electrical and thermal conductivity of pasta, finding spiral defects in molecular dynamics simulations of pasta that could act to scatter electrons. They demonstrate that a signature of the low conductivity pasta layer would be in the thermal behavior of the crust and they show that models of crust cooling in \mxb\ that include a low conductivity pasta layer can account for the observed drop in count rate. Similarly, a low electrical conductivity layer has been suggested by \citet{pons2013} to explain the puzzling cutoff in the spin period distribution of pulsars at $P\sim 10$ seconds, and they suggest the low electrical conductivity layer may be associated with a nuclear pasta phase deep in the crust.



The quasi-free neutrons that coexist with nuclear pasta in the deep inner crust also impact late time crust cooling \citep{page2012}. The critical temperature $T_c$ of the $^1$S$_0$ neutron singlet pairing gap is expected to increase from zero near neutron drip to a maximum value near $T_c \gtrsim 10^9 \, \mathrm{K}$ before decreasing again at high mass densities where the repulsive core of the neutron interaction removes the tendency to form pairs. Calculation of the critical temperature, however, is complicated by the influence of the nuclear clusters, and a wide range of predictions for $T_c(\rho)$ have been made in the literature (e.g., see the plot in Page \& Reddy 2012 and references therein). One of the uncertain aspects of the pairing gap is whether the $^1$S$_0$ gap closes before or after the crust-core transition \citep{chen1993}. If the gap closes before the crust-core transition and there is a low thermal conductivity pasta layer, a layer of normal neutrons will persist near the base of the crust where $T>T_c$, significantly increasing its heat capacity. Here we show that a normal neutron layer with a large heat capacity leaves a signature in the cooling curve at late times and a crust cooling model with normal neutrons gives the best fit to the quiescent cooling observed in \mxb. 


The months to years long flux decays following magnetar outbursts have also been successfully fit with crust thermal relaxation models (e.g.,~\citealt{lyubarsky2002,pons2012,scholz2014}). Many uncertainties remain, including the origin of the X-ray spectrum, the nature of the heating event that drives the outburst, and the role of other heat sources such as magnetospheric currents \citep{Beloborodov2009}. Despite this, magnetar flux decays are interesting because the decay can span a large range of luminosity, and because multiple outbursts from the same source can be studied. The outburst models typically require energy injection into the outer crust of the star, but a significant amount of energy is conducted inward to the core. Late time observations as the magnetar's crust relaxes may then probe the thermal properties of the inner crust. 




We investigate the role of a low thermal conductivity pasta layer and normal neutrons in cooling neutron stars in more detail in this paper. In Section~\ref{sec:model}, we outline our model of the crust cooling in \mxb, highlighting the important role of the density dependence of the neutron superfluid critical temperature near the crust-core transition. In Section 3, we discuss late time cooling in other sources, including the accreting neutron star \ks\ and the magnetar \sgr. We conclude in Section~\ref{sec:discussion}.

\section{The late time cooling of \mxb}
 \label{sec:model}

\subsection{Crust cooling model and the role of the normal neutron layer at the base of the crust}

We follow the thermal evolution of the neutron star crust using the thermal evolution code \dStar \ \citep{dstar} which solves the fully general relativistic heat diffusion equation using a method of lines algorithm in the \mesa \ numerical library \citep{paxton2011,paxton2013,paxton2015}. The microphysics of the crust follows \citet{brown2009}. The results are verified with the code \code{crustcool}\footnote{\code{https://github.com/andrewcumming/crustcool}} which solves the heat diffusion equation assuming constant gravity through the crust. 


We model the $\approx 2.5 \, \mathrm{year}$ outburst in \mxb\ \citep{wijnands2003,wijnands2004} using a local mass accretion rate $\dot{m}  = 0.1\, {\dot{m}_{\rm Edd}}$, where $\dot{m}_{\rm Edd} = 8.8 \times 10^4 \, \mathrm{g \ cm^{-2} \ s^{-1}}$ is the local Eddington mass accretion rate. The model uses a neutron star mass $M = 1.6\, \mathrm{M}_{\odot}$ and radius $R=11.2\,\mathrm{km}$ that are consistent with the \mxb\ quiescent light curve fits from \citet{brown2009}. For the crust composition we use the accreted composition from \citet{haensel2008} that assumes an initial composition of pure $^{56}\mathrm{Fe}$ (see their Table~A3). 

\begin{figure}
\includegraphics[width=\columnwidth]{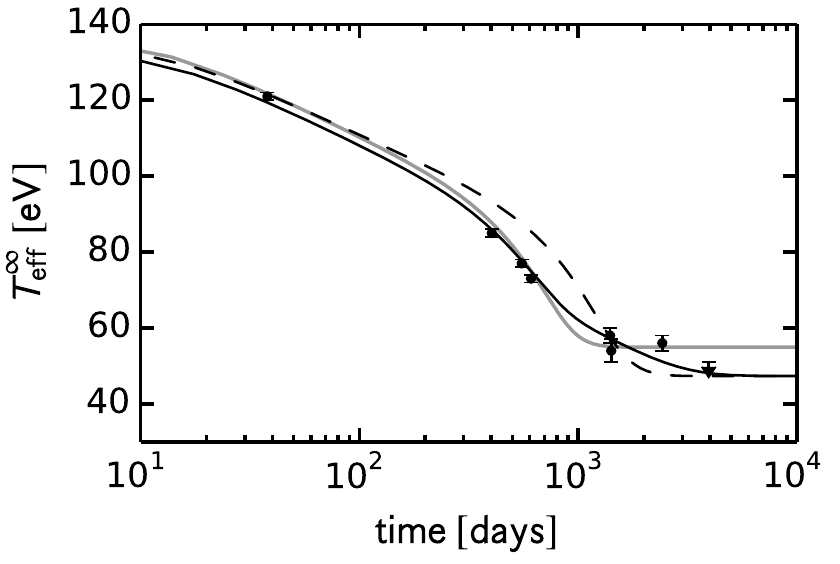}
\caption{\label{fig:figure1}
Cooling models for MXB~1659-29. The solid gray curve is a model that uses $\Qimp=2.5$ throughout the entire crust and $\Tcore = 4 \times 10^7 \, \K$. The solid black curve is a model with $\Qimp=20$ for $\rho>8\times 10^{13}\ {\rm g\ cm^{-3}}$, $\Qimp=1$ for $\rho<8\times 10^{13}\ {\rm g\ cm^{-3}}$, $\Tcore = 3 \times 10^7 \, \K$, and using the \Ggap\ pairing gap. The dashed black curve uses the same $\Qimp$ as the solid black curve, but with the \Sgap\ pairing gap.}
\end{figure}

The thermal conductivity in the inner crust is largely set by impurity scattering. The impurity parameter of the crust is given by
\begin{equation} \label{eq:Qimp}
\Qimp \equiv \frac{1}{n_{\rm ion}} \sum_j n_j \left(Z_j - \Zbar\right)^2 \ ,
\end{equation}

\noindent where $n_{\rm ion}$ is the number density of ions, $n_j$ is the number density of the nuclear species with $Z_j$ number of protons, and $\Zbar$ is the average proton number of the crust composition. The impurity parameter in the neutron star crust was constrained to $\Qimp < 10$ in \mxb \ \citep{brown2009} assuming a constant impurity parameter throughout the entire crust. We show a model of crust cooling in \mxb \ with $\Qimp = 2.5$ and $\Tcore=4 \times 10^7 \, \K$, consistent with the fit from \citet{brown2009}, in Figure~\ref{fig:figure1}. In this model, the crust reaches thermal equilibrium with the core by $\approx 1000$ days into quiescence, and so predicts a constant temperature at later times. 



\begin{figure}
\includegraphics[width=\columnwidth]{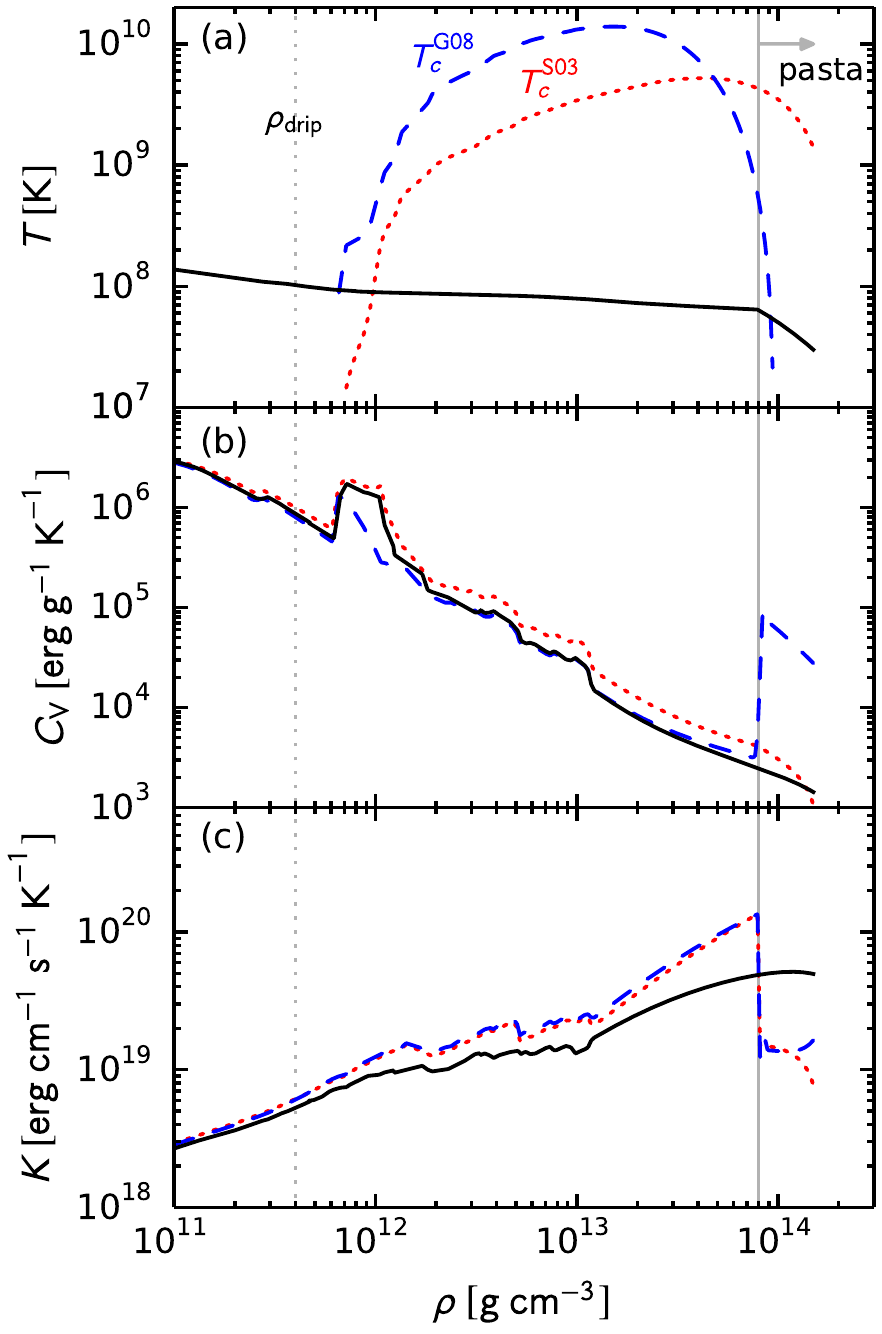}
\caption{\label{fig:figure2} Thermal transport in the inner crust of \mxb\ at the start of quiescence. The gray vertical lines indicate the neutron drip density and the transition to nuclear pasta. {\em Panel (a):} The temperature profile (solid curve) corresponding to the cooling model in Figure~\ref{fig:figure1} with a $Q_{\rm imp}=20$ pasta layer and the \Ggap\ pairing gap. The dashed curves show two choices for $T_c(\rho)$; the blue dashed curve corresponds to \Ggap\ and the red dotted curve is \Sgap. {\em Panel (b):} The heat capacity profiles for the same models as Figure~\ref{fig:figure1}. Solid black curve: $\Qimp=3.7$ throughout the inner crust. Dashed blue curve: $\Qimp=20$ for $\rho>8\times 10^{13}\ {\rm g\ cm^{-3}}$ and $\Qimp=1$ for $\rho<8\times 10^{13}\ {\rm g\ cm^{-3}}$ using the \Ggap\ pairing gap that closes in the crust. Dotted red curve: same as dashed curve, but with a different choice for $T_c(\rho)$ from the \Sgap\ pairing gap that closes in the core. {\em Panel (c):} Thermal conductivity profiles for the same models. } 
\end{figure}

We also run two models with an impure inner crust with $\Qimp=20$ for $\rho>8\times 10^{13}\ {\rm g\ cm^{-3}}$ (and $\Qimp=1$ for $\rho<8\times 10^{13}\ {\rm g\ cm^{-3}}$)  to represent the low conductivity expected for nuclear pasta, as done in \citet{horowitz2015}; both models have a neutron star mass $M=1.6 \, \mathrm{M_{\odot}}$, radius $R=11.2 \, \mathrm{km}$, and $\Tcore = 3 \times 10^7 \, \K$.  The two models use different choices of the neutron superfluid critical temperature profile $T_c(\rho)$. The first uses a $^1$S$_0$ gap that closes in the inner crust (\citealt{gandolfi2008}; hereafter \Ggap) and the second uses a gap that closes in the core (\citealt{schwenk2003}; hereafter \Sgap). The difference in the $T_c(\rho)$ profiles for each pairing gap model are shown in Figure~\ref{fig:figure2}.


As can be seen in Figure~\ref{fig:figure1}, the solid curve corresponding to the \Ggap\ gap shows a long decline in temperature even at late times near $ \approx 4000\, \mathrm{days}$, and thermal equilibrium is reach near $\approx 5000 \, \mathrm{d}$ into quiescence. By contrast, the blue curve using the \Sgap\ gap does not show a decline at late times, but instead levels off to a constant temperature after $\approx 2000 \, \mathrm{days}$. This difference arises because significant late time cooling only occurs if there is a normal layer of neutrons at the base of the crust, giving a large heat capacity there (Fig.~\ref{fig:figure2}). As we show in the following section, a low conductivity pasta layer maintains a temperature difference between the inner crust and core during quiescence and a layer of normal neutrons survives at the base of the crust that has a long thermal time.



\subsection{Analytic estimates} \label{sec:estimates}

Some analytic estimates are useful to understand why the late time cooling occurs, and the crucial role of the normal neutron layer. First, we consider the temperature contrast $\Delta T$ between the inner crust and the core that develops during the accretion outburst. This is set by the value at which the heat flux through the pasta layer balances the nuclear heating in the crust (mostly located at shallower densities near the neutron drip region). The heating rate is $\epsilon_{\rm nuc} = \dot m E_{\rm nuc}$ where $E_{\rm nuc}\approx 2\ {\rm MeV}$ per accreted nucleon. The equivalent heat flux is $F_{\!\rm in} \approx 2\times 10^{22}\ {\rm erg\ cm^{-2}\ s^{-1}}$ for an accretion rate of $\dot m = 0.1\, \dot{m}_{\rm Edd}$. 

The heat flux through the pasta layer is $F\approx K\Delta T/H$, where $K$ is the thermal conductivity and $H$ the pressure scale height. Neutrons set the pressure in the inner crust, so that\footnote{%
In the inner crust, $\Gamma_1\equiv\left(\partial\ln P/\partial\ln\rho\right)_s$ varies with density: at first $\Gamma_1$ decreases below $4/3$ (the value for degenerate relativistic electrons) and then it increases for $\rho\gtrsim 10^{13}\,\mathrm{g\,cm^{-3}}$ and approaches $\Gamma_1\lesssim 2$ at roughly nuclear density. For definiteness in computing $H$, we set $\Gamma_1 = 5/3$.
}
$H=P/\rho g\approx 7\times 10^{4}\ {\rm cm}\ (\rho_{14}^{2/3}Y_n^{5/3}/g_{14})$ 
where $Y_n$ is the neutron fraction, $\rho_{14}$ is the mass density in units of $10^{14} \, \mathrm{g \ cm^{-3}}$, and the surface gravity of the neutron star is $g = (GM/R^2)(1-2GM/Rc^2)^{-1/2}$ in units of $10^{14} \, \mathrm{cm \ s^{-2}}$. The thermal conductivity is primarily set by electron-impurity scattering, with scattering frequency \citep{itoh1993,potekhin99}
\begin{eqnarray}\label{eq:nu_eQ}
\nu_{eQ} &=& {4\pi e^4 n_e\over \pFe^2 \vFe}{\Qimp\over \Zbar} \Lambda_{eQ} \\&\approx& 3\times 10^{18}\ {\rm s^{-1}}\,\left[\left({\rho_{14}Y_e\over 0.05}\right)^{1/3}\left({\Qimp\Lambda_{eQ}\over \Zbar}\right)\right].
\end{eqnarray}
Here $\pFe$ and $\vFe$ and the Fermi momentum and velocity of the electrons, $\Lambda_{eQ}$ is the Coulomb logarithm, and $Y_e$ is the electron fraction. 
The quantity $\Qimp \Lambda_{eQ}/\Zbar$ is of order unity in the inner crust.
The resulting thermal conductivity is
\begin{eqnarray}\label{eq:KeQ}
K_e &=& \frac{\pi}{12} \frac{\EFe \,\kB^2 \, Tc}{e^4} \frac{\Zbar}{\Qimp \Lambda_{eQ}} \nonumber\\ &\approx& 4\times 10^{19}\ {\rm erg \ s^{-1} cm^{-1} K^{-1}}\;\left[ T_8\left({\rho_{14}Y_e\over 0.05}\right)^{1/3}\,\frac{\Zbar}{\Qimp\Lambda_{eQ}} \right],
\end{eqnarray}
where $T_8 \equiv T/(10^8 \, \mathrm{K})$. Therefore, the temperature difference between inner crust and core is 
\begin{equation}
\label{eq:deltaT}
\Delta T \approx 3\times 10^7\ {\rm K}\;\left[{\rho_{14}^{1/3}\over g_{14}T_8}{Y_n^{5/3}\over (Y_e/0.05)^{1/3}}\left({\Qimp \Lambda_{eQ} \over \Zbar}\right)\left({\dot m\over 0.1\dot m_{\rm Edd}}\right)\right] ,
\end{equation}
which is in reasonable agreement with the temperature jumps seen in Figure~\ref{fig:figure1}. 

We can understand the cooling timescale of the normal neutron layer as follows. The specific heat capacity of the normal neutrons is 
\begin{eqnarray}
C_{V} &=&{\pi^2 \over \rho} {n_n \kB^2T\over \pFn \vFn}\nonumber\\&\approx& 3\times 10^{4}\;\mathrm{erg\,g^{-1}\,K^{-1}}\;\left[
Y_{n}^{1/3} \rho_{14}^{-2/3} \left(\frac{T_{7}}{3}\right) 
\right],
\end{eqnarray}
where 
$n_n$ is the number density of free neutrons, $\pFn$ is the neutron Fermi momentum, and $\vFn$ is the neutron Fermi velocity. The thermal diffusivity is
\begin{eqnarray}
D &=& \frac{K}{\rho C_V} \nonumber\\
 &\approx&  4\ {\rm cm^2\ s^{-1}}\ \left({Y_e\over 0.05 Y_n}\right)^{1/3} \left(\frac{\Zbar}{\Qimp \Lambda_{eQ}}\right) ;
\label{eq:D}
\end{eqnarray}
$D$ is independent of temperature and depends only weakly on density. The thermal timescale is then
\begin{eqnarray}
t_{\rm therm} &\approx& \frac{H^2}{D} \nonumber\\
 &\approx& 4000\,{\rm d}
 \left[\rho_{14}^{4/3}Y_n^{11/3}\left({ g_{14} \over 2}\right)^{-2}\left({\Qimp\Lambda_{eQ}\over \Zbar}\right)\left({Y_e \over 0.05}\right)^{-1/3}\right]. 
\end{eqnarray}
Again, this is in good agreement with the cooling timescale we see in the numerical models. It also highlights the role of the large heat capacity from the normal neutrons. Without the normal neutrons, the electrons would set the heat capacity in the inner crust (see, e.g., Fig.~6 of \citealt{brown2009}); in this case, the thermal conductivity is $K=\rho C_V^e \vFe^2/3\nu_{eQ}$ where $C_V^e$ is the heat capacity of electrons, and we see that the thermal diffusivity is then $D\approx c^2/3\nu_{eQ}$  which is about two orders of magnitude larger than that given by Equation~(\ref{eq:D}). Without normal neutrons, the inner crust cools in months, so that no late time cooling signature of the pasta region is seen.

\subsection{Effect of thermal conduction by neutrons} 


The large heat capacity of the normal neutrons suggests they may have a large thermal conductivity that could contribute significantly to the thermal conductivity near the base of the crust. Heat conduction by normal neutrons has been considered in the neutron star core (e.g.,~\citealt{baiko2001}), but not in the crust. We calculate the scattering frequency for neutrons scattering from nuclei in the inner crust, either from thermal vibrations (phonons) or irregularities in the structure (impurities). The details are given in the Appendix; the total scattering frequency is given in Equation ~(\ref{eq:nu_n_tot}). The neutron thermal conductivity is then $K= \pi^2n_n \kB^2 T/(3 m_n^{\star} \nu)$, or 
\begin{eqnarray}
K_n &=& {9\pi^3\over 4}{n_n\kB^2T\over \mnstar}{\hbar \over \mnstar c^2}{n_n\over n_{\rm ion}} \left({\hbar c\over V_0 R_A}\right)^2 \left[\Lambda_{n,{\rm phn}} +\frac{\Qimp}{{\Zbar}^2}\Lambda_{nQ} \right]^{-1}
\nonumber\\
&\approx&3\times 10^{17}\ {\rm erg \ s^{-1} cm^{-1} K^{-1}}\ T_8Y_n\rho_{14}\left({\mnstar \over m_n}\right)^{-2} \left({n_n / n_{\rm ion} \over 100\,}\right) \nonumber\\&& \times \left({\hbar c \over V_0 R_A}\right)^{2}  \left[\Lambda_{n,{\rm phn}} +\frac{\Qimp}{{\Zbar}^2}\Lambda_{nQ} \right]^{-1}.
\end{eqnarray}
In this expression $m_n^{\star}=\pFn ~\left[\partial \varepsilon(p)/\partial p\right]_{p=\pFn}^{-1}$
is the Landau effective mass and $\varepsilon(p)$ is the neutron single particle energy including the rest mass \citep[see, e.g.,][]{Baym1976}. 
The dimensionless quantity $V_0R_A/\hbar c$ is of order unity and measures the strength of the neutron--nucleus interaction; $R_A$ is the typical size of the scattering structure and energy $V_0$ is the magnitude of the scattering potential. Note that the neutron scattering frequency can be approximately reproduced by making the substitutions $e^2\rightarrow V_0R_A$ and $\pFe^2\vFe\rightarrow \pFn^2 \vFn$ in equation (\ref{eq:nu_eQ}) for the electron--impurity scattering rate. The quantities $\Lambda_{n,{\rm phn}}$ and $\Lambda_{nQ}$ are the Coulomb logarithms for phonon and impurity scattering, respectively. As we discuss in the Appendix, we are able to write the impurity scattering for neutrons in terms of the impurity parameter for electron-impurity scattering $\Qimp$.

\begin{figure}
\includegraphics[width=\columnwidth]{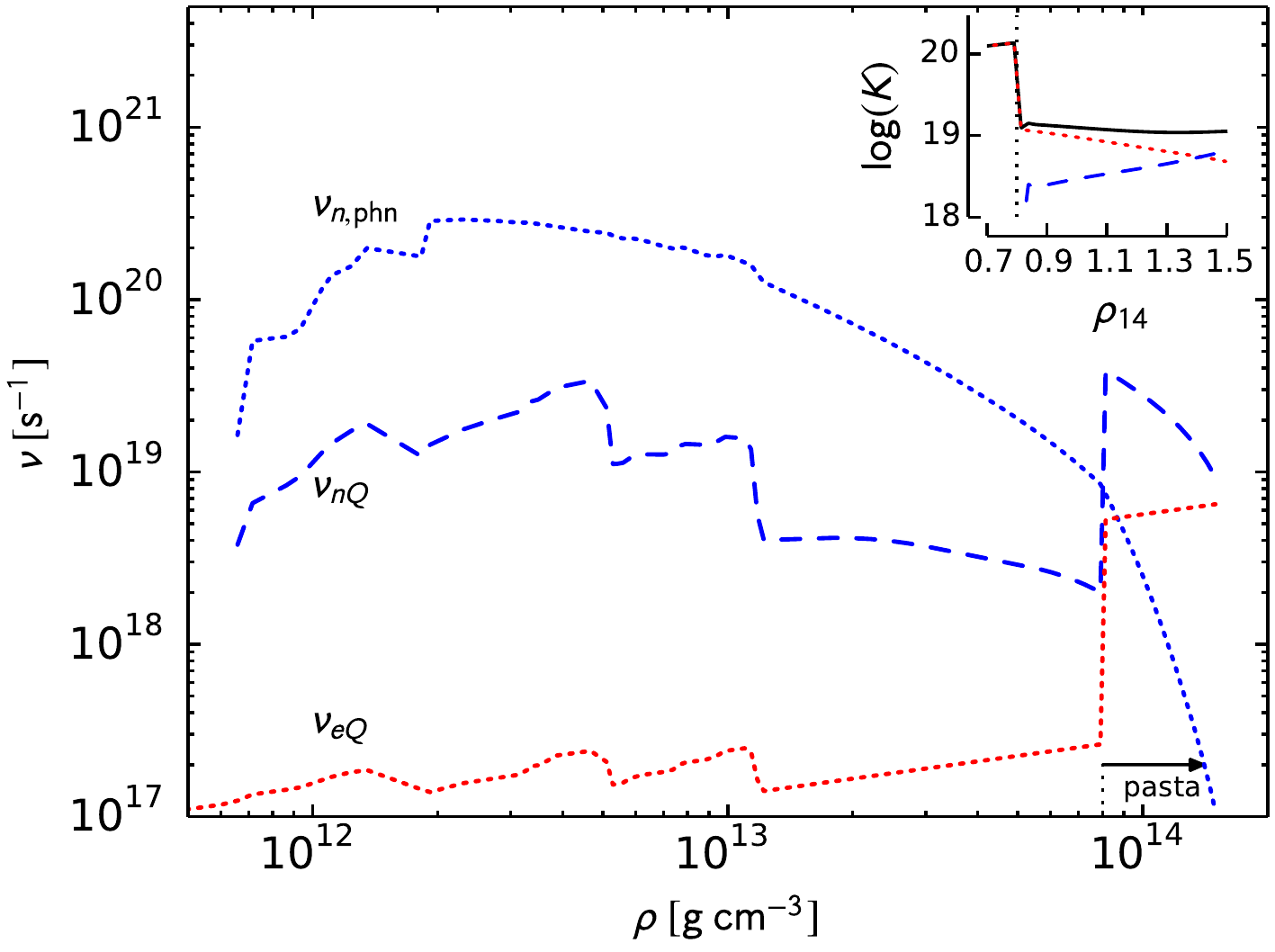}
\vspace{-0.3cm}
\caption{Scattering frequencies in the inner crust at the beginning of quiescence for the model with $\Qimp = 20$ at $\rho > 8 \times 10^{13} \, \mathrm{g \ cm^{-3}}$, $\Qimp = 1$ at $\rho < 8 \times 10^{13} \, \mathrm{g \ cm^{-3}}$, and the pairing gap that closes in the crust \citep{gandolfi2008}. {\em Subplot:} Thermal conductivity $K$ from electron scattering (dotted red curve), neutron scattering (dashed blue curve), and from both electrons and neutrons (solid black curve). The mass density $\rho$ is given in units of $10^{14} \, \mathrm{g \ cm^{-3}}$.  The region containing nuclear pasta is to the right of the vertical black dotted line.
\label{fig:crust_kappa}}
\end{figure}

The thermal conductivity of electrons and neutrons is compared in Figure \ref{fig:crust_kappa} for the crust temperature profile in Figure~\ref{fig:figure2} where we show the separate contributions from phonon and impurity scattering as a function of density. As has been discussed previously, impurity scattering dominates phonon scattering for electrons in the inner crust when $\Qimp\gtrsim 1$ (e.g.,~\citealt{brown2009}). We find for neutron scattering that the phonon contribution is larger where $\Qimp\ =1$ and the impurity contribution is larger in the pasta layer where $\Qimp\ = 20$, as can be seen in Figure~\ref{fig:crust_kappa} (see also Equation~\ref{eq:nu_n_ratio}). 

Figure \ref{fig:crust_kappa} shows that the conductivity due to neutrons can be comparable to the electron conductivity near the base of the crust, but is otherwise not important. To see this in more detail, it is useful to calculate the ratio $K_n/K_e$. The electron thermal conductivity is given by equation (\ref{eq:KeQ}) and taking the impurity contribution to the neutron conductivity only (since $\nu_{n,{\rm phn}}\lesssim \nu_{nQ}$ at the base of the crust), we find
\begin{equation}\label{eq:KnKe}
{K_n\over K_e} = {9 \alpha^2 {\Zbar}^{2/3}}\left({\hbar c\over V_0 R_A}\right)^2{\Lambda_{eQ}\over \Lambda_{nQ}}\left({n_n\over n_{\rm ion}}\right)^{4/3}\left({\vFn\over c}\right)^2,
\end{equation}
where $\vFn = \pFn/\mnstar$ is the Fermi velocity of the neutrons and $\alpha=e^2/\hbar c\approx 1/137$ is the fine structure constant. All the factors in equation (\ref{eq:KnKe}) are $<1$, except for $n_n/n_{\rm ion}$ which is typically $\sim 100$. Therefore, we expect $K_n \lesssim K_e$, consistent with our numerical evaluation shown in Figure \ref{fig:crust_kappa}. Also, since $\vFn = \pFn/m_n^{\star}$, we see that $K_n/K_e$ increases with density approximately as $\rho^{2/3}$, so that neutron thermal conductivity is most important at higher densities. Therefore, we do not expect that neutron thermal conductivity will remove the late time cooling effect of an impure pasta layer.

We emphasize, however, that our calculation of the neutron scattering rates could be improved. The neutron-phonon Umklapp processes dominate because $\kFn a_{\rm ion} \gg 1$, where $a_{\rm ion}=(3/4\pi n_{\rm ion})^{1/3}$ is the inter-ion spacing and $\kFn$ is the neutron Fermi wave number, and the inelastic contribution from the lattice structure function typically denoted as $\delta S_\kappa(q)$ plays a very important role in enhancing neutron-phonon scattering. This suggests that a more detailed analysis of neutron band structure effects is warranted since this suppression at low temperature can be quite significant. Even if neutron-phonon processes are suppressed due to neutron band structure, neutron-impurity scattering will ensure that $K_n \lesssim K_e$ except perhaps in the densest layers. In the highest density regions where pasta phases likely exist the neutron density contrast and therefore $V_0$ are reduced, which makes neutron scattering less efficient. A more detailed calculation including a model of the neutron distribution in the pasta region is needed to take this into account. 

An additional piece of physics that could affect late time cooling is neutrino emission from the pasta layer \citep{leinson1993,gusakov2004a,newton2013}. To compete with the inwards flux of $F_{\rm in}\approx 2\times 10^{22}\ {\rm erg\ cm^{-2}\ s^{-1}}$ (see Section~\ref{sec:estimates}), the neutrino emissivity would have to be $\epsilon_\nu\approx \rho F_{\rm in}/y\sim 10^{18}\ {\rm erg\ cm^{-3}\ s^{-1}}$ at $T\sim (3\textrm{--}6)\times 10^7\ {\rm K}$. Neutrino cooling in the pasta layer can be enhanced when the neutrons are in the normal phase. Two mechanisms for such enhancement have been considered earlier. In one scenario, enhanced neutrino pair emission arises from spin flip transitions of neutrons due to their spin-orbit interactions with the density gradients in the pasta phase \citep{leinson1993}. In this case, estimates indicate that the neutrino emissivity $\epsilon^{\rm pasta}_{\nu\bar{\nu}} \approx 4 \times 10^{23} T^6_9 \, \mathrm{erg \ cm^{-3} \ s^{-1}}$. In the second scenario, the direct Urca processes $e^-+p \rightarrow n + \nu_e$ and $ n\rightarrow e^- + p +\bar{\nu}_e$ are kinematically allowed due to coherent Bragg scattering of nucleons from the the pasta \citep{gusakov2004a}; the resulting neutrino emissivity is $\epsilon^{\rm pasta}_{\rm Urca} \approx 4 \times Y^{1/3}_{e} \times 10^{21} T^6_9 \, \mathrm{erg \ cm^{-3} \ s^{-1}}$. From these estimates and the preceding discussion we conclude that neutrino cooling in the pasta layers, even with normal neutrons, is unlikely to be relevant at the temperatures encountered during thermal relaxation of the neutron stars and magnetars studied here.       

\section{Late time cooling in other sources}

\subsection{Quasi-persistent transients}

The accreting neutron star \ks\ also has quiescent cooling measurements at late times. \cite{merritt2016} recently reported a new temperature measurement for \ks\ taken $\approx 5300\, \mathrm{days}$ into quiescence. They found that the temperature was consistent with the previous value measured $\approx 3000\, \mathrm{days}$ into quiescence, implying that the neutron star crust has now reached thermal equilibrium with the neutron star core near $\Tcore \approx 9.3 \times 10^7 \, \K$. Furthermore, the cooling curve could be fit equally well with or without an impure pasta layer at the base of the crust. This is a similar result to that found by \citet{horowitz2015}, finding that they could fit \ks\ equally well with or without an impure layer. The role of the normal neutrons, however, has not been examined in the late time cooling in this source.

\begin{figure}
\includegraphics[width=\columnwidth]{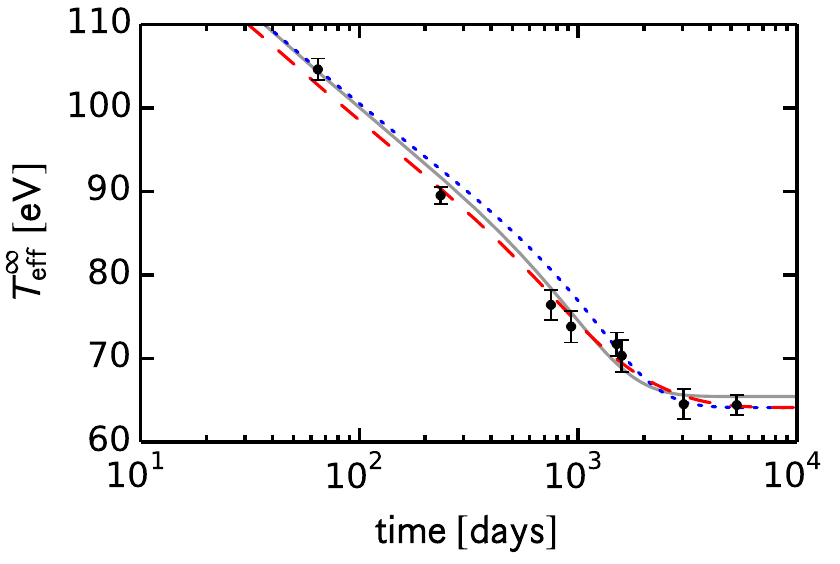}
\caption{\label{fig:ks1731}
Cooling models for \ks. Solid gray curve: Light curve fit from \citet{merritt2016} with: $T_c(\rho)$ from \citet{schwenk2003}, a crust impurity parameter of $Q_{\rm imp} = 4.4$, $T_{\rm core} = 9.35 \times 10^7 \, \mathrm{K}$, and no low conductivity pasta layer. Blue dotted curve: a crust impurity parameter of $\Qimp=2$, a pasta impurity of $Q_{\rm imp}=20$, with $T_c(\rho)$ from \citet{schwenk2003} and $T_{\rm core} = 9.1 \times 10^7 \, \mathrm{K}$. Red dashed curve: a crust impurity parameter of $\Qimp=2$, a pasta impurity of $Q_{\rm imp}=20$, with $T_c(\rho)$ from \citet{gandolfi2008} and $T_{\rm core} = 9.1 \times 10^7 \, \mathrm{K}$.}
\end{figure}

We now model the quiescent cooling of \ks\ following its $\approx 12.5 \, \mathrm{yr}$ outburst including an impure pasta layer and the \Ggap\ gap that closes in the crust. The model uses a neutron star mass $M=1.4 \, \mathrm{M_{\odot}}$ and radius $R=10\,\mathrm{km}$, consistent with the spectral fits and crust models from \citet{merritt2016}. The model uses an iron envelope and an accretion rate $\dot{m} = 0.1 \, \dot{m}_{\rm Edd}$ as done in \citet{merritt2016,cumming2016}. The model fits to the quiescent cooling of \ks\ can be seen in Figure~\ref{fig:ks1731}.

Although the cooling of \ks\ can be fit well without a low conductivity pasta layer \citep{merritt2016}, the cooling model with a $Q_{\rm imp} =20$ pasta layer and the \Ggap\ gap provides a better fit to the data. The fit requires a lower core temperature near $T_{\rm core} \approx 9.1 \times 10^7 \, \K$. The high core temperature means that there is not a large temperature difference $\Delta T$ (see Equation~\ref{eq:deltaT}) between the inner crust and the core during quiescence. The normal neutrons, however, still release heat at late times and the crust reaches thermal equilibrium with the core near $\approx 5000 \, \mathrm{d}$ into quiescence. 


\begin{figure*}
\centering
\includegraphics[width=0.8\textwidth]{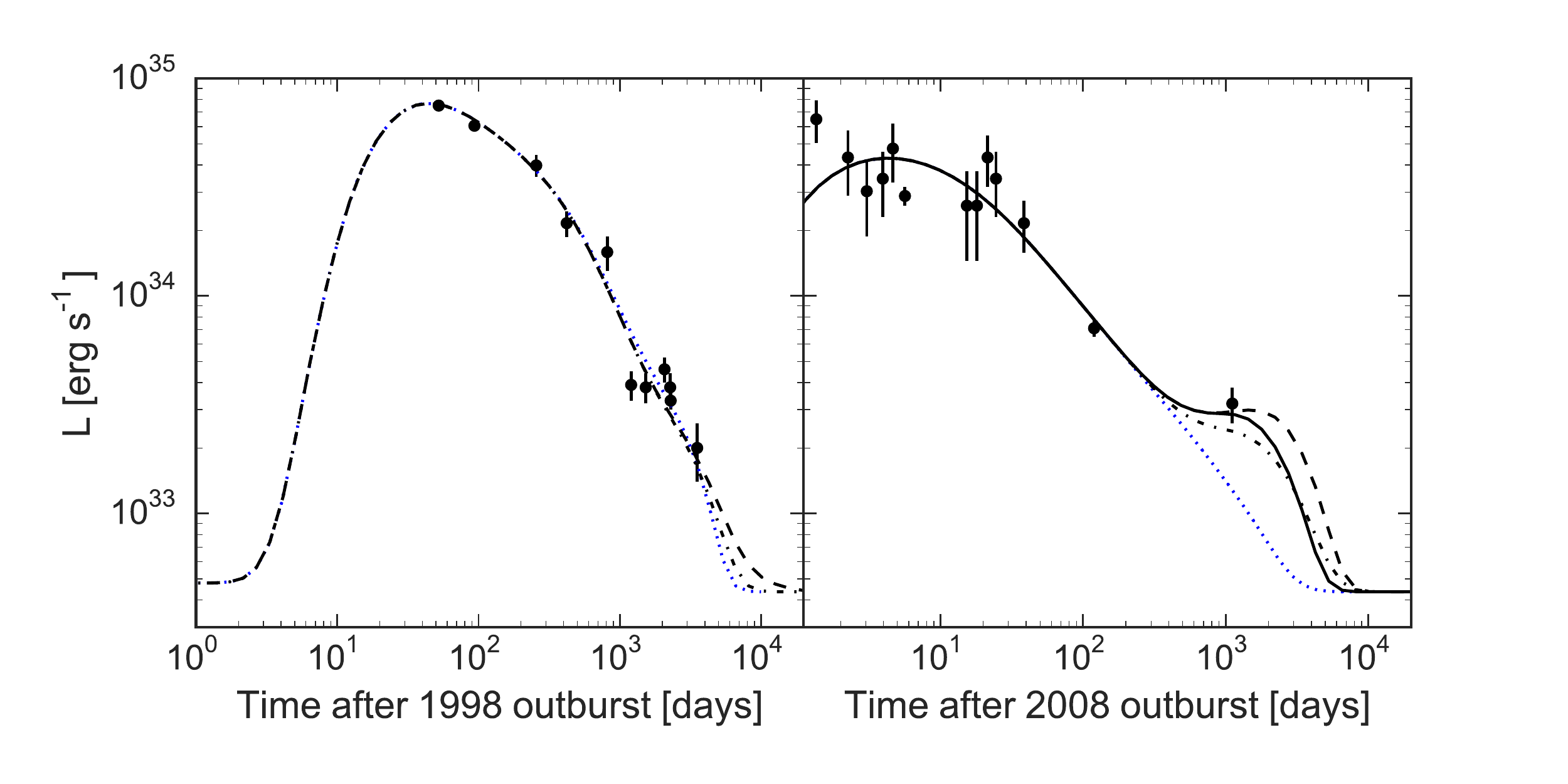}
\caption{\label{fig:sgr}
Cooling models for the 1998 outburst (left) and 2008 outburst (right) of \sgr. The dotted curve in each panel shows the best-fitting model with a constant energy density deposited throughout the crust. The solid curve in the right panel has the same energy density as the 1998 outburst deposited in the inner crust. The dashed curve has an impure pasta layer with $\Qimp=25$ for $\rho>8\times 10^{13}\ {\rm g\ cm^{-3}}$. The dot-dashed curve is for $T_c(\rho)$ case B1 from Page \& Reddy (2011); all other models have $T_c(\rho)$ from Schwenk et al. (2003).}
\end{figure*}

\subsection{The magnetar \sgr}

The magnetar \sgr\ has had two outbursts, one in 1998 \citep{woods1999} and one in 2008 \citep{an2012}. \citet{an2012} showed that the flux decay after the outburst could be reproduced by depositing energy in the outer crust, although with about an order of magnitude difference in the depth and magnitude of the energy deposited ($\sim 10^{43}\ {\rm ergs}$ at $\rho\lesssim 10^{11}\ {\rm g\ cm^{-3}}$ for 1998; $\sim 10^{42}\ {\rm ergs}$ at $\rho\lesssim 10^{10}\ {\rm g\ cm^{-3}}$ for 2008). The measured luminosities are shown in Figure \ref{fig:sgr}. 

The late time observations of \sgr\ after its 1998 outburst are reminiscent of the drop in flux seen in \mxb. The luminosity appeared to have leveled off after $\approx 1000\, \mathrm{d}$, but then showed a drop by a factor of two in an observation at $\approx 3500\, \mathrm{d}$. Moreover, \cite{an2012} showed that the flux $\approx 1000\, \mathrm{d}$ after the 2008 outburst was similar to the flux at the same time after the 1998 outburst. This is unexpected because the energy deposition in the 2008 outburst is much shallower, and by $\approx 1000\, \mathrm{d}$ the crust should have relaxed to the core temperature and the luminosity should be at its minimum value. \cite{an2012} suggested that perhaps the last flux measurement after the 1998 outburst was a statistical deviation (it is within 2$\sigma$ of the previous flux value), and that the luminosity seen at $\approx 1000\, \mathrm{d}$ in both outbursts reflects the core temperature. This would mean \sgr\ then has a hot core with a temperature near $T_{\rm core} \approx 10^8\ {\K}$. 

Here, we pursue the possibility of a colder core and investigate whether the drop at $\approx 3000\, \mathrm{d}$ days is due to inner crust physics. We model the flux decay of \sgr\ using \code{crustcool}, which includes envelope models and thermal conductivities that take into account the strong magnetic field (averaged over angle around the star; \citealt{scholz2014}). The dotted curves in Figure \ref{fig:sgr} show the best fitting model with a constant energy density deposited throughout the crust (we find similar values as \citealt{an2012}), and a low value of $Q_{\rm imp}=3$ throughout the crust. We set the core temperature to $T_{\rm core} = 7\times 10^7\, {\K}$ (as measured at the crust-core boundary) so that the flux in the 1998 outburst continues to decline at $\approx 3000\, \mathrm{d}$. As \cite{an2012} pointed out, the 2008 outburst then cools much too quickly to agree with the flux measured at $\approx 1000\, \mathrm{d}$.

We introduce a pasta layer with $Q_{\rm imp}=25$ at $\rho > 8 \times 10^{13} \, \mathrm{g \ cm^{-3}}$ to try to increase the luminosity at $\approx 1000\, \mathrm{d}$ after the 2008 outburst. Unlike accreting neutron stars, however, we find that introducing a low conductivity layer in the pasta region is not enough to delay the cooling. The difference is that in the accreting case the heating is over a long timescale so that the temperature of the inner crust is increased substantially. In the magnetar case, the energy is deposited at the beginning of the outburst and the high temperature of the inner crust must be established rapidly. We find that with our colder core temperature, the only way to get agreement with the $\approx 1000\, \mathrm{d}$ 2008 outburst measurement is to deposit extra energy in the inner crust at the beginning. The solid curve in Figure \ref{fig:sgr} shows a model that matches both outbursts, in which we deposit the same energy density in the inner crust in the 2008 outburst as in the 1998 outburst (but keep different amounts of energy in the outer crust).

If the core temperature is low $T_{\rm core}\lesssim 7 \times 10^7 \, \K$ in \sgr, future observations should show a further decline in flux. We find that the future evolution is sensitive to the choice of $\Qimp$ in the pasta layer and the choice of $T_c(\rho)$. This is shown by the dashed, solid and dot-dashed curves in Figure~\ref{fig:sgr} that have different choices for those parameters (these models are all consistent with the 1998 outburst, left panel of Fig.~\ref{fig:sgr}).

\section{Discussion} 
\label{sec:discussion}


We have examined the late time quiescent cooling of the neutron star transient \mxb\ where cooling was observed for $\approx 4000\,\mathrm{d}$ into quiescence prior to its renewed outburst activity \citep{negoro2015}. The quiescent cooling probes successively deeper layers of the neutron star's crust with increasingly longer thermal times and the late time cooling $\gtrsim 1000\, \mathrm{days}$ into quiescence depends on the thermal transport properties of the inner crust. In particular, late time cooling in \mxb\ requires a low thermal conductivity layer with $\Qimp\ \gtrsim 20$ at mass densities $\rho \gtrsim 8 \times 10^{13} \, \mathrm{g \ cm^{-3}}$ where nuclear pasta is expected to appear \citep{horowitz2015}. The pasta layer maintains a temperature difference of $\Delta T \approx 3 \times 10^7 \, \K$ between the inner crust and core during the outburst. As a consequence, normal neutrons with a long thermal time appear at the base of the crust that cause late time cooling if the neutron singlet pairing gap closes in the crust. Without normal neutrons at the base of the crust, as is the case if the neutron singlet pairing gap closes in the core, the crust reaches thermal equilibrium with the core after $\approx 3000 \, \mathrm{d}$ and late time cooling is removed. 






Page \& Reddy (2012) pointed out that differences in $T_c(\rho)$ and the resulting presence or absence of a layer of normal neutrons at the base of the crust could affect the cooling curves at late times $\approx 1000$ days into cooling. We find a much larger effect and on a longer timescale here because the low thermal conductivity of the nuclear pasta layer keeps the inner crust much hotter during the outburst. During quiescence, the base of the crust remains at a higher temperature than the core for $\approx 5000 \, \mathrm{days}$ (see Equation~[\ref{eq:deltaT}]). The temperature difference between the crust and core results in a slow decline of the quiescent light curve after $\gtrsim 1000 \, \mathrm{days}$, as can be seen in Figure~\ref{fig:figure1}. 

We also investigated the late time cooling of \ks\ which was observed $\approx 14.5 \, \mathrm{yrs}$ into quiescence \citep{cackett2013}. Although the quiescent light curve in this source can be fit without a low conductivity pasta layer \citep{merritt2016}, we find a comparable fit with a $\Qimp = 20$ pasta layer and using a neutron singlet pairing gap that closes in the crust (see Figure~\ref{fig:ks1731}). That both \mxb\ and \ks\ fits prefer a pasta layer with $\Qimp=20$ suggests that the inner crust composition may be similar in accreting neutron stars regardless of their initial crust composition, as was found in a study of the accreted multi-component crust \citep{gupta2008}. 

We studied \sgr, a magnetar with late time observations of two outbursts. Based on the previous outburst in 2008, the source may not yet have fully thermally relaxed and could show further cooling. We investigated a low conductivity pasta region as a way to prolong the cooling, but found that the flattening of the luminosity at times $\gtrsim 1000$ days could be explained only if energy was deposited directly into the inner crust. This is because in magnetars the energy is assumed to be deposited rapidly rather than over many thermal times as in accreting neutron stars. Nevertheless, if the core temperature is low $T_{\rm core}\lesssim 7 \times 10^7 \, \K$, variations in inner crust physics affect the light curve and should be included in models. Furthermore, the need for energy to be deposited in the inner crust constrains models for transient magnetic energy release in magnetars (e.g.,~\citealt{Li2016,Thompson2016}), and argues against only heating the crust externally (e.g.,~\citealt{li2015}).



Late time cooling in \mxb\ requires that the $^1$S$_0$ neutron singlet pairing gap close in the crust. As a result, superfluid neutrons are confined to the inner crust shallower than the pasta layer at $\rho \lesssim 8 \times 10^{13} \, \mathrm{g \ cm^{-3}}$ where $T \ll T_c$. By contrast, a recent study of pulsar glitches suggests that the neutron superfluid extends from the crust into the core continuously \citep{andersson2012}. Recent calculations of the neutron effective mass in a non-accreted ($\Qimp = 0$) crust suggest that $\mnstar \gg m_n$ at the base of the crust \citep{chamel2005,chamel2012}. In this case, a larger fraction of free neutrons are entrained in the inner crust and the neutron superfluid must then extend into the core to supply adequate inertia for pulsar glitches \citep{andersson2012}. We note, however, that the above calculation for the neutron effective mass is likely inappropriate for the impure crust compositions found in the accreting transients studied here. Therefore, we here assume $\mnstar \approx m_n$ as found in \citet{brown2013} in the absence of effective mass calculations in an accreted crust. 

\acknowledgements

Support for A. D. and E. F. B. was provided by the National Aeronautics and Space Administration through Chandra Award Number TM5-16003X issued by the Chandra X-ray Observatory Center, which is operated by the Smithsonian Astrophysical Observatory for and on behalf of the National Aeronautics and Space Administration under contract NAS8-03060. A. D. and E. F. B. are also supported by the National Science Foundation under Grant No.\ AST-1516969. A. D. is also grateful for the support of the Michigan State University College of Natural Science Dissertation Completion Fellowship. A.C. is supported by an NSERC Discovery grant, is a member of the Centre de Recherche en Astrophysique du Qu\'ebec (CRAQ), and an Associate of the CIFAR Cosmology and Gravity program. S. R. was supported by the U.S. Department of Energy under Contract No.\ DE-FG0200ER41132. This material is based upon work supported by the National Science Foundation under Grant No.\ PHY-1430152 (JINA Center for the Evolution of the Elements).

\appendix

\section{Neutron scattering frequency in the inner crust}
\label{sec:appendix}

In this Appendix, we derive expressions for the neutron scattering frequency in the inner crust. In the relaxation time approximation, the scattering frequency can be expressed as \citep{flowers1976,potekhin99}
\begin{equation} \label{eq:nui}
\nu_n =  \frac{\mnstar}{12\pi^3 \hbar^3} \frac{n_{\rm ion}}{n_n} \int_0^{2 \kFn} dq\, q^3 \, |V(q)|^2 \, S_{\kappa} (q) \,,\end{equation}
where $\hbar q$ is the momentum transfer, $\pFn=\hbar (3 \pi^2 n_n)^{1/3} \equiv \hbar\kFn$ is the neutron Fermi momentum, $\mnstar$ is the effective neutron mass, $n_n$ is the number density of neutrons outside of nuclei, $n_{\rm ion}$ is the number density of ions, and $V(q)$ is the the Fourier transform of the scattering potential.  The scattering medium is described by the structure function
\begin{equation} \label{eq:skappa}
S_{\kappa}(q) = \int_{- \infty}^{\infty} \frac{d \omega}{2 \pi} \frac{\hbar \omega}{\kB T} \frac{S(q,\omega)}{1-\exp(-\hbar \omega/\kB T)}  \left[1 +  \left(\frac{\hbar \omega}{\kB T} \right)^2  \left(3 {\kFn^2\over q^2} - {1\over 2} \right)\right],
\end{equation}
which is written in terms of the dynamical structure factor $S(q,\omega)$. 

To describe neutron-nucleus scattering in the inner crust, we assume that the nuclei are spherical and that the surface thickness is negligible compared to the size of the nucleus. Although the nuclei in the  pasta phase are certainly non-spherical, a description of scattering in non-spherical geometries is beyond the scope of this paper. Under these assumptions, the potential seen by the neutrons can be modeled as a square well with $V(r<R_A) = V_0$, where $R_A$ is the radius of the scattering center. The depth of the potential $V_0\approx V_{\rm in} - V_{\rm out}$, where $V_{\rm in}$ and $V_{\rm out}$ are the neutron single particle potentials inside and outside the scattering structures, respectively. In the pasta phase, the density contrast between the scattering structure and the background rapidly decreases with increasing density, implying a correspondingly rapid decrease in $V_0$ and reduced neutron scattering. 

With the spherical assumption, the effective neutron-nucleus potential in momentum space is
\begin{equation} \label{eq:Vna}
V_{n,A}(q) = V_0 \frac{4 \pi R_A^3}{3} F_A (q R_A)\ ,
\end{equation}
with a form factor \citep{flowers1976}
\begin{equation}
F_A (x) = \frac{3[\mathrm{sin}(x) - x\, \mathrm{cos}(x)]}{x^3} \ .
\end{equation}
The form factor $F_A \rightarrow 1$ in the limit that momentum transfers are small ($x = qR_A \ll 1$) and is suppressed when momentum transfers are large. 
Inserting equation~(\ref{eq:Vna}) into equation~(\ref{eq:nui}), we find the neutron-phonon scattering frequency
\begin{equation}
\nu_{n,{\rm phn}} = \frac{4}{27 \pi} \frac{\mnstar c^2}{\hbar} {n_{\rm ion}\over n_n} \left(\frac{V_0 R_A}{\hbar c}\right)^2 \Lambda_{n,{\rm phn}} \ ,
\end{equation}
where the Coulomb logarithm is given by
\begin{equation}\label{eq:lambda_ph}
\Lambda_{n,{\rm phn}} = \int_0^{2 \kFn R_A} dx\, x^3 \, F_A^2(x) \, S_{\kappa}^{\rm phn} (q=x/R_A) \ .
\end{equation}
We evaluate the integral in equation (\ref{eq:lambda_ph}) using a Runge-Kutta scheme of order 8(5,3) \citep{Hairer1993} and fitting formulae for $S^{\rm phn}_\kappa(q)$ \citep[][Equations~21 and 22]{potekhin99} that were developed in the context of electron-phonon scattering.

We find the frequency of neutron-impurity scattering using a similar approach. We assume that the impurities are uncorrelated elastic scatterers, and write the scattering frequency as a sum over all impurity species. The neutron-impurity potential for an impurity of species $j$ with radius $R_j$ is
\begin{equation} \label{eq:Vjn}
V_{n,j} = V_0 \frac{4 \pi \bar{R}^3}{3} F_A(q\bar{R}) \left(\frac{R_j^3}{\bar{R}^3}\frac{F_A(qR_j)}{F_A(q\bar{R})} -1 \right) \ ,
\end{equation}
where $\bar{R}$ is the radius of the average ion in the lattice.  With this assumption, the dynamical structure factor for impurity scattering is \citep{flowers1976}
\begin{equation}\label{eq:dynamicalS-imp}
S^{\rm imp} (q, \omega) = {1\over n_{\rm ion}} \sum_j 2 \pi n_j \delta(\omega) \ ,
\end{equation}
where $j$ is the sum over impurity species and $n_j$ is the number density of impurities. 

Upon using equations~(\ref{eq:dynamicalS-imp}) and (\ref{eq:skappa}) to obtain 
$S_\kappa^{\mathrm{imp}} (q) = \sum_j n_j/n_{\mathrm{ion}}$, 
and inserting $S_\kappa^{\mathrm{imp}}(q)$ and $V_{n,j}$ (eq.~[\ref{eq:Vjn}]) into equation~(\ref{eq:nui}), we find the neutron-impurity scattering frequency
\begin{equation}
\nu_{nQ} =  \frac{4}{27 \pi} \frac{\mnstar c^2}{\hbar} \frac{n_{\rm ion}}{n_n} \left(\frac{V_0 \bar{R}}{\hbar c}\right)^2 {\Lambda}_{nQ} \, \widetilde{Q}. 
\end{equation}
Here we define the Coulomb logarithm for neutron-impurity scattering,
\begin{equation}
\Lambda_{nQ} = \int^{2 \kFn \bar{R}}_0 dx \, x^3\, F_A^2(x) , 
\end{equation}
and the impurity parameter for neutron scattering,
\begin{equation}
\widetilde{Q} = \frac{1}{\Lambda_{nQ}} \int^{2 \kFn\bar{R}}_0 dx \, x^3\, F_A^2(x) \, \sum_j \frac{n_j}{n_{\rm ion}} \left(\frac{R_j^3}{\bar{R}^3}\frac{F_A(xR_j/\bar{R})}{F_A(x)} -1 \right)^2.
\end{equation}
For scattering involving momentum transfers $q\lesssim 1/R_j$ the ratio 
$F_A(qR_j)/F_A(q\bar{R})\approx 1$. Taking $R_j^3\propto Z_j$ and $\bar{R}^3 \propto \Zbar$ then gives $\widetilde{Q}\approx \Qimp/\Zbar^2$ where $\Qimp$ (Eq.~[\ref{eq:Qimp}]) is the impurity parameter for electron scattering. The neutron-impurity scattering frequency is therefore
\begin{equation}
\nu_{nQ} \approx \frac{4}{27\pi} \frac{\mnstar c^2}{\hbar} \frac{n_{\rm ion}}{n_n}\left(\frac{V_0 \bar{R}}{\hbar c}\right)^2\,\frac{\Qimp}{{\Zbar}^2}\Lambda_{nQ} \ .
\end{equation}

Since the neutron chemical potentials inside and outside the nucleus are required to be equal in Gibbs equilibrium, we can estimate $V_0$ as the difference in the single particle kinetic energies inside and outside the nucleus,
\begin{equation} 
V_0\approx    \frac{\hbar^2(3\pi^2n_{\rm in})^{2/3}}{2m_n}~\left( 1-\left( \frac{n_{n}}{n_{\rm in}}\right)^{2/3}\right) ,
\end{equation}
where $n_{\rm in}$ is the neutron number density inside the nucleus. We take $R_A$ to be the proton radius of the nucleus given by $(4\pi/3)R_A^3n_{\rm in}=Z$, where $Z$ is the proton number of the nucleus.
We therefore expect that $V_0R_A/\hbar c \sim \mathcal{O}(1)$ in the inner crust. 
The total scattering frequency is
\begin{eqnarray}
\nu_n = \nu_{n,{\rm phn}}+\nu_{nQ} 
	& \approx & \frac{4}{27\pi} \frac{\mnstar c^2}{\hbar} 
    	\frac{n_{\rm ion}}{n_n}
    	\left(\frac{V_0R_A}{\hbar c}\right)^2 
        \left[ \Lambda_{n,{\rm phn}} + \frac{\Qimp}{\Zbar^2}\Lambda_{nQ}\right]\nonumber\\ 
    &=& 6.7\times 10^{20}\,\mathrm{s^{-1}}\;\left(\frac{\mnstar}{m_n}\right) 
    	\left(\frac{n_{\rm ion}/n_n}{0.01}\right)
        \left(\frac{V_0 R_A}{\hbar c}\right)^2 
        \left[ \Lambda_{n,{\rm phn}} + \frac{\Qimp}{{\Zbar}^2}\Lambda_{nQ}\right].
\label{eq:nu_n_tot}
\end{eqnarray}
The ratio of the phonon and impurity scattering frequencies is
\begin{equation}\label{eq:nu_n_ratio}
{\nu_{n,{\rm phn}}\over \nu_{nQ}}=\frac{{\Zbar}^2}{\Qimp}{\Lambda_{n,{\rm phn}}\over\Lambda_{nQ}}
\end{equation}
and is typically of order unity for $\Qimp \simeq 10$.

\bibliographystyle{apj}
\bibliography{ms}

\end{document}